\documentclass{sf2a-conf2018}
\usepackage{graphicx}
\usepackage{hyperref}
\usepackage[]{natbib}  
\usepackage{epstopdf}

\def\BibTeX{{\rm B\kern-.05em{\sc i\kern-.025em b}\kern-.08em
    T\kern-.1667em\lower.7ex\hbox{E}\kern-.125emX}}
\bibpunct{(}{)}{;}{a}{}{,}  

\begin{document}

\TitreGlobal{SF2A 2018}


\title{Massive stars with Pollux on LUVOIR}

\runningtitle{Massive stars with Pollux}

\author{C. Neiner}\address{LESIA, Paris Observatory, PSL University, CNRS, Sorbonne Université, Univ. Paris Diderot, Sorbonne Paris Cité, 5 place Jules Janssen, 92195 Meudon, France}

\author{J.-C. Bouret}\address{Aix Marseille Université, CNRS, LAM (Laboratoire d'Astrophysique de Marseille) UMR 7326, F-13388 Marseille, France}

\author{C. Evans}\address{UK ATC, Royal Observatory, Blackford Hill, Edinburgh, EH9 3HJ, UK}

\author{the Pollux Hot Stars working group}

\setcounter{page}{237}


\maketitle


\begin{abstract}
Many open questions remain about massive stars, for example about their evolution, their wind, and their maximum mass at formation. These issues could be ideally adressed by the Pollux UV spectropolarimeter onboard LUVOIR. Here we present examples of the science themes that one could study with Pollux regarding massive stars. 
\end{abstract}



\section{Pollux onboard LUVOIR}

Pollux is a high-resolution (R=120000) spectropolarimeter working in the ultraviolet (UV) domain. It is studied by a European consortium for the LUVOIR 15-m space telescope project proposed to the NASA Decadal 2020 survey. Pollux covers the wavelength range from 90 to 400 nm in 3 arms: the far UV (FUV) spectrum is observed separately, while the mid UV (MUV) and near UV (NUV) domains are observed simultaneously. Each of the 3 spectrographs (FUV, MUV, and NUV) is equiped with its own dedicated polarimeter. More details about this instrument can be found in Bouret et al. (these proceedings).

\section{Massive stars}

Massive stars provide heavy chemical elements to the Universe and dominate the interstellar radiation field. Moreover, they are the progenitors of supernovae, neutron stars, black holes, gamma-ray bursts, and gravitational waves. In addition, due to their luminosity and spectroscopic features, the successive phases of massive stars and starbursts can be observed out to large distances. Therefore, they are essential for many domains of astrophysics, such as stellar and planetary formation and galactic structure and evolution. 

About $\sim$10\% of massive stars host a magnetic field of fossil origin, usually dipolar but inclined with respect to the stellar rotation axis, with a polar field strength ranging from a few hundreds to a few thousands Gauss \citep{neiner2015, grunhut2015}. The $\sim$90\% of stars that do not host such a field may nevertheless host an ultra-weak field of the order of 1 Gauss, such as those recently discovered in some A and Am stars \citep[e.g.][]{blazere2016}. The presence of a magnetic field, even a weak one, is crucial for stellar structure and evolution and has a strong impact on the circumstellar environment. 

Since massive stars emit most of their radiation in the UV domain, and show atomic and molecular lines coming from the photosphere and wind in this wavelength range, they are ideal targets for a UV spectropolarimeter like Pollux. 

\section{Examples of observing programs with Pollux}

\subsection{Environments}

\begin{figure}[t!]
 \centering
 \includegraphics[width=\textwidth,clip]{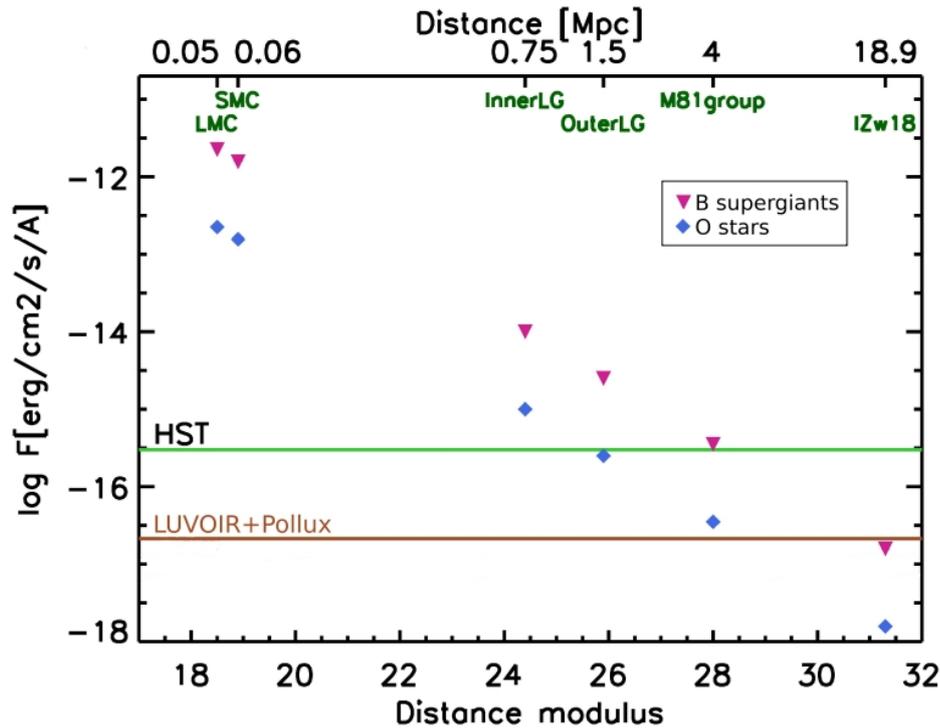}      
  \caption{Flux level observed for B supergiants (pink triangles) and O stars (blue diamonds) as a function of distance compared to the limit of HST (green line) and LUVOIR+Pollux (brown line). Adapted from a figure kindly provided by M. Garcia.}
  \label{distance}
\end{figure}

Thanks to the 15-m primary mirror of LUVOIR and to the high-resolution spectrograph of Pollux, it will be possible to obtain very high quality UV spectra of weak massive stars in various environments. In particular, it will be possible to observe these targets in the Large Magellanic Clouds, in the Small Magellanic Cloud, in the inner and outer Local Group, as well as in M81 (see Fig.~\ref{distance}). This will allow us to test the effect of metallicity on various stellar parameters, e.g. on mass loss and wind. The high-quality UV spectra will also allow us to characterise the late stages of evolution in Local Group galaxies. 

Thanks to the polarimetric module of Pollux, it will also be possible to measure, for the first time, magnetic fields in a large number of stars outside our own galaxy. It will then be possible, e.g., to check the effect of the environment on the presence and properties of the magnetic fields in massive stars. 

\subsection{Stellar wind}

Thanks to the UV high-resolution spectra obtained by Pollux for massive stars, we will be able to accurately measure mass loss rates, wind terminal velocities, and wind variability and clumpiness. Indeed, these parameters are best observed in the UV in the resonance lines sensitive to the wind. With this information at hand, we will be able to study the effect of rotation and metallicity on the stellar wind, the consequences of the wind and mass loss on stellar evolution and on the feedback into the interstellar medium.

\begin{figure}[t!]
 \centering
 \includegraphics[width=0.55\textwidth,clip]{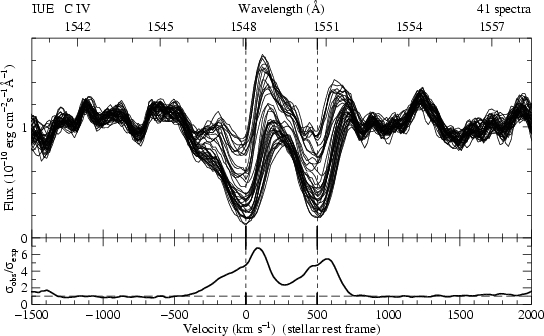}      
 \includegraphics[width=0.44\textwidth,clip]{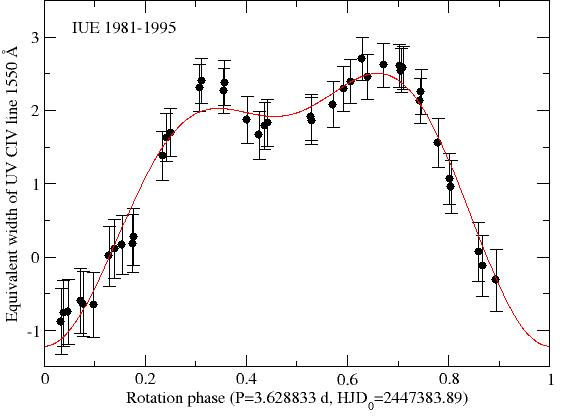}      
  \caption{Left: Variation of the wind-sensitive UV line of C{\sc iv} at 155 nm in the B star V2052\,Oph observed by IUE. Right: Equivalent width of that line, folded in phase with the rotation period. Adapted from \cite{neiner2003}.}
  \label{iue}
\end{figure}

In the case of magnetic massive stars, the stellar wind is magnetised and its particles are channeled along magnetic field lines into a magnetosphere around the star \citep[e.g.][]{owocki2014}. Thanks to the Pollux spectropolarimeter, we will be able to measure the rotational modulation produced by the obliquity between the magnetic axis and the rotation axis. In particular, the magnetised wind emerges from the magnetic poles and we can also observe eclipses due to the magnetosphere located in the magnetic equator plane. This will allow us to determine the rotation period of the star very precisely (see Fig.~\ref{iue}), as was already done for a few stars with the IUE archive \citep[e.g.][]{neiner2003}.

By performing spectropolarimetric measurements directly into wind-sensitive resonance lines in the UV domain, we will also be able to measure for the first time the magnetic field directly in the wind (rather than at the stellar surface as it is done in the visible domain). We will then be able to derive a 3D map of the circumstellar environment and study directly the link between what happens around the star (magnetosphere, co-rotating interaction regions,...) and at its  surface (spots, mass ejections,...).

\subsection{Initial mass function}

One of the open questions about massive stars is whether there is a universal maximum stellar mass and whether the stellar initial mass function is the same in galaxies where the stellar formation rate is much more intense than in the Milky Way \cite[e.g.][]{andrews2013}. Thanks to the 15-m primary mirror of LUVOIR, it will be possible to statistically compare massive star populations in various environments allowing to answer these questions.

\begin{figure}[t!]
 \centering
 \includegraphics[width=\textwidth,clip]{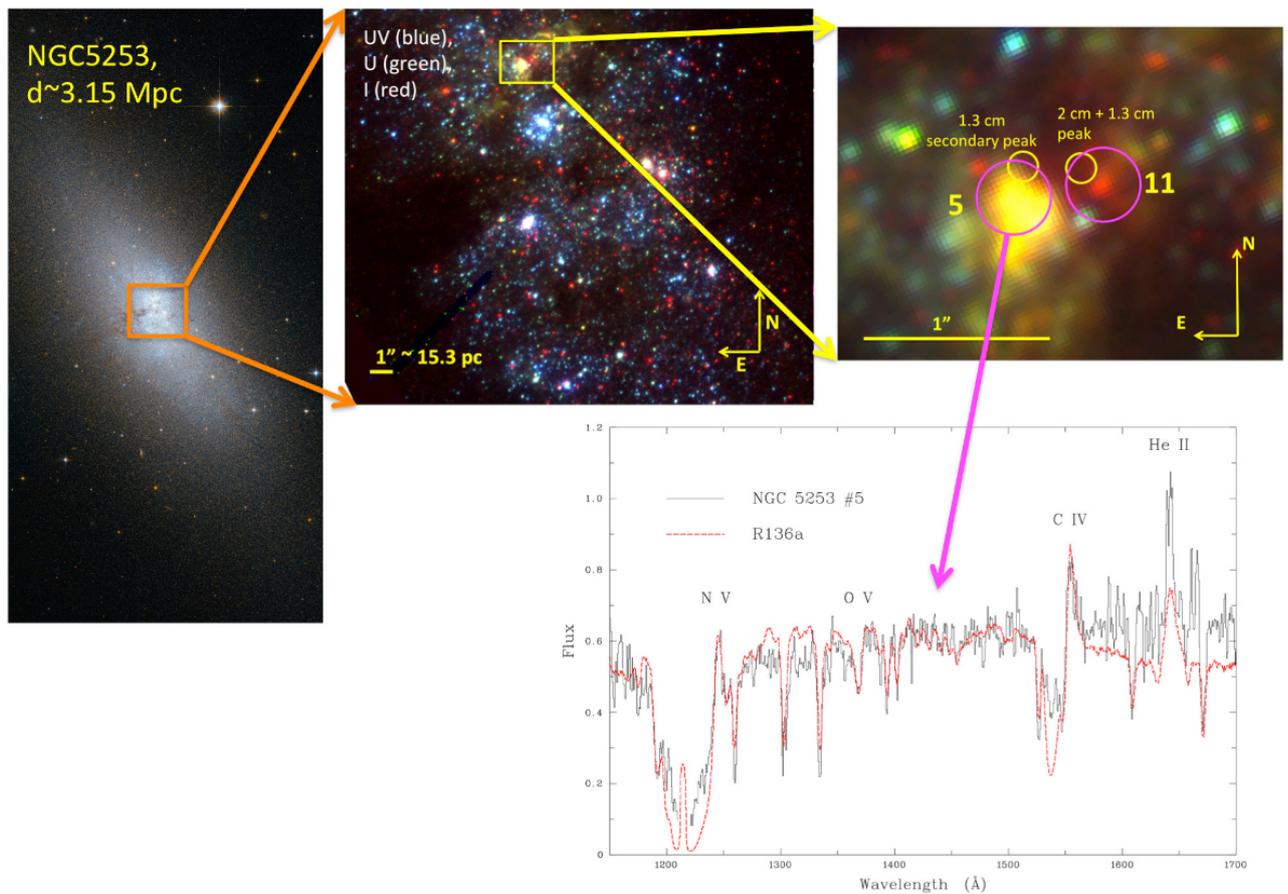}      
  \caption{Observations of the NGC\,5253 cluster \citep[by][]{calzetti2015} containing a very massive star (NGC\,5253 \#5) as confirmed by its UV spectrum obtained with HST (black line in bottom right panel, from \cite{smith2016}). The spectrum is compared to the one of another very massive star, R136a (red line), scaled to the distance of NGC\,5253.}
  \label{vms}
\end{figure}

One may also wonder whether very massive stars really exist. Such targets are difficult to recognize in the visible domain and can easily be confused with Wolf-Rayet stars for example. However, their signature is obvious in the UV domain (see Fig.~\ref{vms}): one expects to observe P Cygni profiles for the N{\sc v} line at 124 nm and C{\sc iv} line at 155 nm, large He{\sc ii} emission at 164 nm, wind absorption in O{\sc v} at 137 nm shifted towards the blue, but no P Cygni for the Si{\sc iv} line at 140 nm. 

\subsection{Multiple systems}

In the Milky Way, massive stars are often found to be binaries or multiple systems \citep{sana2017}. However, we do not know if this is also the case in other environments with other densities, metallicities, ages,... This multiplicity rate is important because binary evolution may impact, e.g., the distribution of rotational velocities in the host galaxy and the production of runaway stars. Moreover, observations of massive stars with Pollux will permit to test predictions of chemically-homogeneous evolution in low metallicity binaries as pathfinders of gravitational wave progenitors. 

In addition, when a binary system host two magnetic stars and when the two stars are close enough to each others, their magnetic field lines may reconnect. This produces a transfer of material and of angular momentum between the two components that can be observed by tracing the wind particles along the field lines in the UV.

\section{Conclusions}

Massive stars are ideal targets for Pollux onboard LUVOIR. High-resolution UV spectroscopy of these objects in various environments would cast light on many open questions in this field of research. In addition, UV spectropolarimetry of massive stars in the UV domain would be performed for the first time and would allow in particular to study the magnetised wind and circumstellar environment.

\begin{acknowledgements}
This work as made used of the SIMBAD database operated at CDS, Strasbourg (France), and of NASA's Astrophysics Data System (ADS). The Pollux instrumental study is supported by CNES, the French Space Agency. 
\end{acknowledgements}

\bibliographystyle{aa}  
\bibliography{sf2a-Neiner-Pollux} 

\end{document}